\documentclass{IEEEtran}
\usepackage[utf8]{inputenc}
\usepackage{amssymb}
\usepackage{amsmath}
\usepackage{bm}
\usepackage{graphics}
\usepackage{hhline}
\usepackage{url}
\usepackage[english]{babel}
\usepackage{graphicx}
\usepackage{amsthm}
\theoremstyle{plain}

\usepackage{stackengine}

\usepackage[table]{xcolor}
\usepackage{multirow}
\title{Empirical Analysis of Indirect Internal Conversions in Cryptocurrency Exchanges }
\newcommand{\eqdef}{\mathrel{\mathop:}=}
\newcommand{\pair}[2]{\left(\text{#1}\mapsto\text{#2}\right)}

\usepackage{amsmath,mleftright}
\usepackage{xparse}

\ifdefined\final
 \newcommand{\authorsnote}[2]{}
 \else
 \newcommand*{\authorsnote}[2]{\textcolor{#1}{[#2]}}
\fi

\NewDocumentCommand{\evalat}{sO{\big}mm}{%
  \IfBooleanTF{#1}
   {\mleft. #3 \mright|_{#4}}
   {#3#2|_{#4}}%
}

\author{

\IEEEauthorblockN{Paz Grimberg, Tobias Lauinger, Damon McCoy} \\
\IEEEauthorblockA{New York University}

}

\begin{document}
\newtheorem*{definition*}{Definition}
\newtheorem{definition}{Definition}[section]
\newtheorem{remark}{Remark}
\newtheorem*{remark*}{Remark}
\newtheorem{exercise}{Exercise}
\newtheorem*{exercise*}{Exercise}
\newtheorem{obsv}{Observation}
\newtheorem*{obsv*}{Observation}

\maketitle
\section{Abstract}

Algorithmic trading is well studied in traditional financial markets. However, it has received less attention in centralized cryptocurrency exchanges. The Commodity Futures Trading Commission (CFTC) attributed the $2010$ flash crash, one of the most turbulent periods in the history of financial markets that saw the Dow Jones Industrial Average lose $9\%$ of its value within minutes, to automated order ``spoofing'' algorithms. In this paper, we build a set of methodologies to characterize and empirically measure different algorithmic trading strategies in Binance, a large centralized cryptocurrency exchange, using a complete data set of historical trades. We find that a sub-strategy of triangular arbitrage is widespread, where bots convert between two coins through an intermediary coin, and obtain a favorable exchange rate compared to the direct one. We measure the profitability of this strategy, characterize its risks, and outline two strategies that algorithmic trading bots use to mitigate their losses. We find that this strategy yields an exchange ratio that is $0.144\%$, or $14.4$ basis points (bps) better than the direct exchange ratio. $2.71\%$ of all trades on Binance are attributable to this strategy.

\section{Introduction}
Cryptocurrency exchanges today handle more than \$50b in daily trade volume \cite{coinmarketcap_daily_volume}. Most of it occurs on \textit{centralized} exchanges, which hold assets and settle trades on behalf of their customers. Traders on these exchanges can convert different coins at certain exchange ratios, just like traditional foreign exchange (FX) markets. The ability to convert between coins creates potential arbitrage opportunities, where traders can make a profit through a series of conversions. The case involving three conversions, coin $c_1$ converted to coin $c_2$, which is then converted to coin $c_3$ and back to $c_1$, is called \textit{triangular arbitrage} if the proceeds of the conversions are greater than the initial quantity. The existence of triangular arbitrage in foreign exchange markets is well documented \cite{tri_arb2} \cite{tri_arb3} \cite{tri_arb1} \cite{Gebarowski2019}. The characteristics of cryptocurrency exchanges and their relationship to traditional foreign exchange markets have been studied as well \cite{Drozdz2019SignaturesOT}. However, to the best of our knowledge, triangular arbitrage has never been studied within the context of a centralized cryptocurrency exchange.

In this paper, we measure arbitrage activity in Binance, a centralized cryptocurrency exchanges, by empirically exploring its complete historical trade data. Since different traders cannot be identified from trade data, we cluster sequences of consecutive trades that match in their quantities and timing and attribute them to the same trader. To the best of our knowledge, we are the firsts to employ a clustering methodology for identifying triangular arbitrage traders, based on trade-by-trade data.

We find that triangular arbitrage is rarely accomplished. Participants predominantly engage in an alternative strategy, which we call \textit{indirect internal conversions}, where coin $A$ is converted to coin $B$ through an intermediary coin $x$, at a favorable exchange ratio compared to directly converting $A$ to $B$. This activity accounts for $2.71$\% of the total daily volume, and offers an exchange ratio that is $14.4$ bps better on average.

We believe that the fee structure in cryptocurrency exchanges makes it unprofitable for participants to engage in triangular arbitrage. Instead, participants turn to indirect conversions as an efficient way to rebalance their holdings. 



\section{Background}
\subsection{Exchanges}
    An exchange is an organized market where tradable securities, commodities, foreign exchange/cryptocurrencies (or ``coins'') and derivatives are sold and bought (collectively referred to as \textit{instruments}). In a centralized exchange, the deposits and assets of participants are held and settled by the exchange. In decentralized exchanges (or ``DEXes''), a smart contract (a program executing on a blockchain) or other form of peer-to-peer network executes exchange functionality. In DEXes, funds cannot be stolen by the exchange operator, because their custody and exchange logic is processed and guaranteed by the smart contract. 
    
    In centralized cryptocurrency exchanges, different cryptocurrencies can be exchanged to others, such as Bitcoin and Ethereum. In addition, some exchanges list ERC20 tokens, or simply ``tokens,'' that can also be exchanged to cryptocurrencies. Tokens are essentially smart contracts that make use of the Ethereum blockchain \cite{ERC20_def}.
    
    Participants can place \textit{orders} on an exchange. An order is
    an instruction to buy or sell some traded instrument. These instructions can be simple or complicated, and can be sent to either a broker or directly to an exchange via direct market access. There are some standard instructions for such orders. For example, a \textit{market order} is a buy or sell order to be executed immediately at the current market prices, i.e., buy at the lowest asking price or sell to the highest bidding price. Market orders are typically used when certainty of execution is a priority over the price of execution. A \textit{limit order} is an order to buy an instrument at no more than a specific price, or to sell at no less than a specific price (called ``or better'' for either direction). This gives the trader control over the price at which the trade is executed; however, the order may never be executed (``filled''). Limit orders are typically used when the trader wishes to control price rather than certainty of execution.
    
    Each instrument traded on an exchange has an \textit{order book}. The order book refers to an electronic list of buy and sell orders for a specific security or financial instrument organized by price level. An order book lists the number of shares being bid on or offered at each price point, or market depth \cite{order_book_def}.
    
\subsection{Arbitrage}
    The traditional financial industry has settled on three main types of quantitative strategies that are sustainable because they provide real economic value to the market: arbitrage, market-making and market-taking \cite{headlands_tech_blog}. In market-taking strategies, traders post both buy and sell orders in the same financial instrument, hoping to make a profit on the bid-ask spread \cite{radcliffe1997}. In market-taking strategies, traders engage in longer-term trading, subject to some rules-based investment methodology. Market-taking strategies on centralized cryptocurrency exchanges have been studied in \cite{krueger2019event}. 
    
    Arbitrage and its economic benefits have been well understood for quite some time and documented by academia \cite{shleifer1997limits} \cite{bjork2009arbitrage}. Competitive arbitrageurs on centralized exchanges have at least one of three advantages.
        \begin{enumerate}
            \item \textit{Scale:} participants who trade large volumes are often compensated in the form of kickbacks, rebates, and low (or zero) trading fees, which provide such participants the opportunity to make profits in cases where others with less scale cannot. 
            \item \textit{Speed:} the ability to access order book information and trade faster than others provides an opportunity to gain from mispricings. For example, triangular arbitrage on FX products is a major impetus for the Go West and Hibernia microwave telecommunications projects \cite{go_west} \cite{hibernia}, where multi-million dollar network infrastructure was developed for the purpose of shaving off milliseconds in electronic trading latency.
            \item \textit{Queue position:} being able to enter one leg of an arbitrage trade by placing orders ahead of time, without crossing the spreads, i.e., placing orders that execute immediately without being queued in the order book, significantly reduces fees, which enables profitable completion of triangular arbitrage trades. Participants are compensated for the risk of queuing orders in the order book. 
        \end{enumerate}
    Arbitrage strategies involving multiple centralized cryptocurrency exchanges, exploiting different prices of same assets, have been studied \cite{kruckeberg2018decentralized} \cite{kroeger2017law} \cite{MAKAROV2020293} \cite{fischer2019statistical}.
    
    Arbitrage strategies on decentralized exchanges are executed by bots who pay high transaction fees and optimize their network latency to front-run ordinary users' trades \cite{flashboys}.
    
    At time of writing, the vast majority of cryptocurrencies trading volume (over $99\%$) is done in centralized exchanges \cite{coinmarketcap_daily_volume}, therefore in this paper, we focus on triangular arbitrage in a single centralized cryptocurrency exchange (Binance). Triangular arbitrage is an arbitrage strategy resulting from a discrepancy between three coins that occurs when their exchange rates give rise to a profitable sequence of trades, i.e., trades of the form $c_1 \mapsto c_2 \mapsto c_3 \mapsto c_1$, where $c_1, c_2, c_3$ are coins, and the ending balance of $c_1$ is greater than the initial balance. We call $c_1$ the \textit{base coin}, $c_2$ and $c_3$ \textit{intermediary coins}, and the sequence a \textit{cycle} \cite{tri_arb_def}. 
    
    To the best of our knowledge, we are the firsts to employ a clustering methodology for identifying triangular arbitrage traders.

\section{Dataset}
\begin{figure}[t]
  \centering
    \includegraphics[width=.50\textwidth]{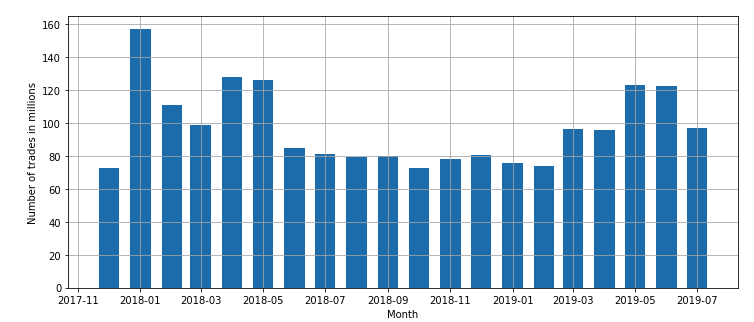}
    \caption{Number of monthly trades, in millions, executed on Binance, based on Kaiko's data set.}\label{daily_num_trades}
\end{figure}

We use historical cryptocurrency trade data and order book data provided by Kaiko from Nov $9$th, $2017$ through Aug $6$th, $2019$ \cite{kaiko}. We conducted our measurements on Binance's exchange data, as it is the largest centralized exchange by daily volume and has the highest number of listed \textit{symbols} (the ``ticker'' of an exchange pair). Kaiko's data consists of trade-by-trade data and market depth data for all cryptocurrencies and ERC20 tokens traded on Binance.  There are $1,964,461,269$ trades. Figure~\ref{daily_num_trades} shows the monthly number of trades executed on Binance.

Every coin in Binance is denominated by at least one of the following coins: BNB (Binance's native token), Bitcoin, ALTS (Etheruem, Ripple, Tron) and stable coins. Stable coins are fiat-pegged cryptocurrencies designed to minimize the volatility of price fluctuations.  We call these coins \textit{anchor coins}, as these coins can be directly exchanged to many other coins and have historically had the greatest market value (in US dollar terms) \cite{binance_markets}.

There are $204$ different coins and $704$ different possible conversions. We denote such conversions by $c_1 \Leftrightarrow c_2$, where $c_1$ and $c_2$ are two different coins. We call $c_1 \Leftrightarrow c_2$, a \textit{pair}. For example, if $c_1=\text{BTC}$ and $c_2$ = \text{ETH}, then the pair is denoted $\text{BTC} \Leftrightarrow \text{ETH}$. Traders can exchange their BTC to ETH (or vice versa) using this conversion.  We write $c_1 \mapsto c_2$ when considering the side of the $c_1$ holder.

There are 6,534 possible cycles in Binance, i.e., sequences of the form $c_1 \mapsto c_2 \mapsto c_3 \mapsto c_1$. In $3,900$ of them, one of $c_2$ and $c_3$ is an anchor. In $2,634$ of them, both $c_2$ and $c_3$ are anchors. There are no cases where both $c_2$ and $c_3$ are non-anchors. $c_1$ can be anchor or non-anchor.

Cycle statistics in Binance are summarized in Table~\ref{cycle_stats}. There are $1,982$ cycles with a non-anchor coin as the base coin. These cycles have the potential to create arbitrage gains in non-anchor coins. However, anchor coins represent over $92\%$ of the total market value of all listed coins and historically had lower volatility, compared to non-anchor coins. Therefore, we focus on cycles with anchor base coins. As future work, we could explore cycles with no-anchor base coins. However, we note that there is inherent risk in operating in non-anchor coins, due to volatility.

\begin{table}[t]
\begin{center}
    \begin{tabular}{||c|c||}
    \hline \text{Base coin} & \text{Number of cycles} \\\hline
    \text{BTC} & 986 \\ \hline
    \text{BNB} & 818 \\ \hline
    \text{ALTS} & 908 \\ \hline
    \text{Stable coins} & 1,840 \\ \hline
    \text{Other coins} & 1,982 \\ \hline
    \text{Total} & 6,534 \\ \hline
\end{tabular}
\end{center}
\caption{Number of cycles in Binance by base coin. BTC,BNB,ALTS and stable coins represent over $92\%$ of the total market value of all listed coins and historically had lower volatility. Therefore, we omit other cycles in this study.}
    \label{cycle_stats}
\end{table}

\subsection{Data Fields}
Kaiko's historical trade data fields are described in Table~\ref{kaiko_trade_data_fields}. The granularity of the \textit{date} field is in ms. Since multiple trades can execute within the same ms, we use the monotonically increasing trade ids as a way to order the trades.

\begin{table}[t]
\begin{center}
    \begin{tabular}{||c|c||}
    \hline \text{Field} & \text{Description} \\\hline
    \text{id} & \text{Incremental trade id} \\\hline
    \text{exchange} &  \text{Exchange id} \\ \hline
    \text{symbol} & \text{Pair symbol} \\ \hline
    \text{date} & \text{Time (in epoch) when the trade took place} \\ \hline
    \text{price} & \text{Trade price} \\ \hline
    \text{amount} & \text{Trade quantity (quoted in base asset)} \\ \hline
    \text{sell} & \begin{tabular}{@{}c@{}}Binary. If TRUE, then the \\ holder of $c_1$, swapping to $c_2$, \\ was a liquidity taker, and $c_1$ is the \\ base asset in the pair \\ listed on the exchange \end{tabular} \\ \hline
\end{tabular}
\end{center}
\caption{Kaiko's trade data fields.}
    \label{kaiko_trade_data_fields}
\end{table}

Kaiko's historical order book data is given on a 60-second basis, that is, for every pair listed, its entire order book is given in 60 second increments of the \textit{date} field. Note that the quotes' depth are not given and need to be reconstructed separately. The data fields are described in Table~\ref{kaiko_order_book_data_fields}.

\begin{table}[ht]
\begin{center}
    \begin{tabular}{||c|c||}
    \hline \text{Field} & \text{Description} \\\hline
    \text{symbol} & \text{Pair symbol} \\ \hline
    \text{date} & \text{Time (in epoch) when the order was placed} \\ \hline
    \text{type} & \text{Bid or ask} \\ \hline
    \text{amount} & \text{Order quantity (quoted in base asset)} \\ \hline
\end{tabular}
\caption{Kaiko's order book data fields.}
    \label{kaiko_order_book_data_fields}
\end{center}
\end{table}

\subsection{Data Limitations}
Kaiko's order book data is given on a $60$-second interval basis and Binance's API does not provide historical order-book data \cite{binance_api}. Complete order book data could reveal all arbitrage opportunities, not just the ones executed with actual trades, as it would accurately paint the traders' view of the market at any point in time. When merging Kaiko's historical trade data with its order book, only about $5\%$ of the trade data has matching order book data available within a $100ms$ interval. (During our measurements, we used smaller time intervals than $100ms$.) For profitability measurements, about $0.5\%$ of the time, we were able to use the order book's bid/ask directly. For the remainder, we approximated the bid/ask price with the volume-weighted average price of that time interval, based on the trade data. In addition, Kaiko does not include trade fees or historical fee schedules, so we had to reconstruct Binance's historical fee time series.

\subsection{Binance Trading Fees}
An important data point for our analysis are the trading fees collected by Binance \cite{binance_fees}. For every trade executed, Binance collects a fee from the proceeds of the trade. The fees are collected either in the currency of the proceeds or in Binance's native ERC20 token, BNB, depending on the trader's BNB balance. If the trader has enough BNB balance, Binance will calculate the conversion rate between the proceeds and BNB, and withdraw the fees from the BNB balance. In general, Binance rewards participants who trade large volumes by charging them a lower fee. In addition, for high-volume traders, Binance distinguishes between liquidity-taking trades, i.e., market order trades or limit order trades that were matched immediately at the best bid/ask level, and market-making trades, which are trades that were placed somewhere in the order book but were not executed immediately. Binance charges less fees for market-making trades, as they wish to encourage participants to ``fill up'' the order book,  thereby narrowing the bid/ask spread and increasing liqudity. This is a common practice; however, in traditional financial exchanges, market-makers pay zero or even negative fees (rebates, kickbacks).

We assume that arbitrageurs are operating at the lowest fee level. To track Binance's fee schedule, we used the \text{Wayback Machine} \cite{wayback}, a digital archive of the World Wide Web, to view Binance's fee web page historically. In our analysis time span, Binance's fee web page changed $46$ times. However, the lowest fee level remained constant at $1.2$bps for makers and $2.4$bps for takers.

\section{Methodology}
In this section, we describe our methodology for discovering triangular arbitrage trade sequences in Binance by examining the trade data. During our analysis, we discovered that a sub-strategy of triangular arbitrage, involving only the first two conversions, is widely deployed. We refined our original methodology to identify such conversions. These conversions exhibit some risks, one of them is the scenario where multiple traders compete for the same intermediary coin, potentially harming laggards' profitability. We identify these risks and outline a methodology for clustering competing events. Lastly, we observe different strategies traders take to mitigate this risk and describe a methodology to detect it.

\subsection{Discovering Triangular Arbitrage Trading Sequences}
To discern triangular arbitrage trading sequences, we design a methodology to identify likely arbitrage trading sequences. We look for trading sequences of the form $c_1 \mapsto c_2 \mapsto c_3 \mapsto c_1$, where $c_1$ is an anchor coin (BTC, BNB, ALTS, Stable coins). Arbitrageurs start with some quantity $Q_1$ of $c_1$, exchange it to $c_2$ at price $p_{12}$, resulting in $Q_2$ units of $c_2$. In practice, Binance deducts the fee upon conversion, therefore $Q_2$ will be slightly lower than the conversion price. Next, $Q_2$ units of $c_2$ are converted to $c_3$, resulting in $Q_3$ units, minus fees. Lastly, $Q_3$ units of $c_3$ are converted back to $c_1$, minus fees, resulting in $Q_1'$ units of $c_1$. If $Q_1'>Q_1$, then the arbitrage sequence is profitable. 

To successfully profit from this opportunity, arbitrageurs need to execute $3$ trades, $c_1 \mapsto c_2$, $c_2 \mapsto c_3$ and $c_3 \mapsto c_1$. First, an arbitrageur needs to calculate the initial quantity $Q_1$ of $c_1$, such that during conversions, fee payments and other trading constraints will leave no or minimal residue in the intermediary coins. Second, since Binance does not support batch trading, i.e., grouping multiple orders in a single request, arbitrageurs need to ensure correct timing of their trades. For example, if the order $c_2 \mapsto c_3$ arrives before $c_1 \mapsto c_2$, then it will fail as the arbitrageur does not hold $c_2$ yet. Furthermore, the arbitrageur competes with other arbitrageurs for these trades, so speed is an important factor. 
 
In order to identify triangular arbitrage sequences, we first need to ensure that the same quantity is flowing through different coins, ending at the base coin. Binance quotes quantities based on how the pair is listed. Different pairs have different quantity/price restrictions, namely, minimum, maximum and increment size. Since quantities and prices change in discrete steps, a small quantity might not be converted, leaving a residue. These small residues need to be accounted for when identifying trades with equal quantities. While these residues have a (small) value, in practice, they cannot be converted immediately to anchor coins, due to minimum size restrictions. As residue accumulates beyond the exchange threshold, arbitrageurs can convert it back to anchor coins. We found that less than $0.01\%$ of the time, the profitability was decided by the residue. Therefore, in our detection process we ignore the residue value. To illustrate coin exchanges, we follow an actual trade example from Binance.

\subsubsection{BTC$\Leftrightarrow$ETH Exchange Example}
Consider the conversion between BTC and ETH. It is listed as ETH/BTC in Binance. ETH is called \textit{base asset} and BTC is called the \textit{quote asset}. The quantity is always given in base asset terms. This pair has a $0.001$ minimum quantity (base asset), $100,000$ maximum quantity, $0.001$ quantity increments, $0.000001$ minimum price (quote asset), $100,000$ maximum price and $10^{-6}$ price increments.  At the time of writing, the last trade of ETH/BTC was executed at a price of $0.018876$ and quantity $0.153$. The ``sell'' flag was on, meaning that the holder of ETH was the one to initiate the trade and met the buyer at the buyer's bid price. Assume both participants are at the cheapest fee level, currently at $2.4$bps for takers and $1.2$bps for makers.

From the ETH holder's perspective: Since ETH is the base asset and the ``sell'' flag is on, this means the ETH holder initiated the trade and met the buyer at the buyer's bid price, therefore paying a liquidity taking fee of $2.4$bps. The holder exchanged $0.153$ units of his ETH at a price of $0.018876$ BTC per $1$ ETH. This means the holder ended with $0.153\cdot0.018876\cdot0.99976=0.002887334873$ units of BTC. Note that if $0.002887334873$ BTC are to be exchanged for some other coin, only $0.00288733$ could be converted, leaving $0.000000004873$ BTC residue.

From the BTC holder's perspective: Since BTC is the quote asset and the ``sell'' flag is on, this means the BTC holder provided the liquidity for the trade and his price was met by the seller, thus paying a liquidity making fee of $1.2$bps. The BTC holder exchanged $0.153\cdot0.018876=0.002888028$ units of BTC at a price of $0.018876$ BTC per $1$ ETH, while paying $1.2$bps fee, resulting in $(0.002888028/0.018876)\cdot0.99988 = 0.15298164$ ETH.

From Binance's perspective: They collect $2.4$bps of the BTC proceeds from the ETH holder, i.e., $0.00024\cdot0.002888028 \approx 6.93\cdot10^{-7}$ units of BTC. If the ETH holder also holds BNB, then this amount is actually converted to BNB terms (using an average of recent exchange ratios between BTC and BNB), and deducted from the BNB balance. If the ETH holder does not own BNB, then $6.93\cdot 10^{-7}$ BTC are deducted from the proceeds. In addition, Binance also collect $1.2$bps of the ETH proceeds from the BTC holder, i.e., $0.153\cdot0.00012 \approx 18.36\cdot 10^{-6}$ units of ETH. If the BTC holder also holds BNB, this amount is collected in BNB terms as well. Note that if the seller/buyer did not hold BNB, the fee would have been higher. In our analysis, we assumed arbitrageurs operate at the lowest fee level.

\subsubsection{Identifying Equal Quantities in Triangular Arbitrage Sequences}
We identify sequences of trades, $c_1 \mapsto c_2 \mapsto c_3 \mapsto c_1$, where the same quantity is passed. The quantity is quoted in the base asset and depends on whether $c_1/c_2$ or $c_2/c_1$ is listed, whether $c_2/c_3$ or $c_3/c_2$ is listed and whether $c_1/c_3$ or $c_3/c_1$ is listed. When matching quantities between trades of two pairs, having a common coin, $c_1 \mapsto c_2$ and $c_2 \mapsto c_3$, we translate the quantities to the common coin's terms, $c_2$, and check if they are equal, up to fees and residue resulting from the trade.  In Table \ref{tab:equality}, we describe the translation process, based on different listing scenarios in Binance.

\begin{table}[t]
\begin{center}
    \begin{tabular}{||c | c||} 
    \hline \text{Binance listing} & \text{Matching condition}  \\\hline
    $c_2/c_3$, $c_1/c_2$ & $q_{12}p_{12} - q_{23} = \text{fee + residue}$ \\ \hline
    $c_2/c_3$, $c_2/c_1$ & $q_{12} - q_{23} = \text{fee + residue}$ \\ \hline
    $c_3/c_2$, $c_1/c_2$ & $q_{12}p_{12} - q_{23}p_{23} = \text{fee + residue}$ \\ \hline
    $c_3/c_2$, $c_2/c_1$ & $q_{12} - q_{23}p_{23} = \text{fee + residue} $ \\ \hline
\end{tabular}
\end{center}
 \caption{Matching equal quantities in conversions with common coin. $q_{ij}$ and $p_{ij}$ are the quantities and prices quoted in the trade data. }
    \label{tab:equality}
\end{table}

\subsubsection{Identifying Trade Latency in Triangular Arbitrage Sequences} When a triangular arbitrage opportunity presents itself, arbitrageurs compete amongst each other to be the first to execute the trade sequences. There are three trades to be executed, $c_1 \mapsto c_2$, $c_2 \mapsto c_3$ and $c_3 \mapsto c_1$. Since Binance does not support batch trading, an arbitrageur will have to send three limit orders with prices $p_{12},p_{23},p_{31}$ and quantities $q_{12},q_{23},q_{31}$. These quantities are equal in the conversion process, in the sense explained previously and the prices match the prices quoted in the order book. To ensure that trades are received in their original order, the arbitrageur will take into account network latency and wait a small amount of time between sending consecutive orders. Analysis of the trade data suggests that the average latency between consecutive trades is $20$ ms with a standard deviation of $15$ ms. Around $95\%$ of trades with matching quantities are executed within $10$ms-$30$ms of each other. Therefore, clustering trades using longer latency does not have material impact on the number of trades identified. We find that using $\Delta t \approx 50 $ms as an approximation for arbitrage latency between consecutive trades maximizes precision and minimizes recall.

We found that less than $0.01\%$ of triangular arbitrage sequences contain a liquidity-making trade. We believe this behavior is caused by Binance's fee model, which charges traders a commission even for liquidity making trades. If traders used liquidity-making orders, they would need to pay the fee in case they were filled. At that point, it is not guaranteed that an arbitrage opportunity will exist, while the fee is already paid and exposure to what is mostly an illiquid coin is already established. To further enhance our precision, we require arbitrage trade sequences to be all liquidity-taking. 

Furthermore, we found a surprisingly low number of triangular arbitrage trades, 1,381,928 in total, accounting for $0.24\%$ of total number of trades. However, in the course of clustering triangular arbitrage trades, we witnessed a much larger number of partial arbitrage sequences, 20,892,825, or $2.71\%$ of total trades, where traders executed the first two trades, $c_1 \mapsto c_2$ and $c_2 \mapsto c_3$, but did not execute the third trade, $c_3 \mapsto c_1$. Executing the first two trades effectively converts $c_1$ to $c_3$ at an exchange ratio of $p_{12}\cdot p_{23}$, minus fees and residue. Interestingly, $95\%$ of the time these trades resulted in an exchange ratio that was favorable to the exchange ratio of the direct trade $c_1 \mapsto c_3$. We call this trading strategy \textit{indirect internal conversions} and explain below in more details what is means to have a ``favorable'' rate. 

We refine our methodology to identify indirect internal conversions and study the root cause of unprofitable instances. We elaborate on the risks associated with this trading strategy in the discussion section.

\subsection{Discovering Indirect Internal Conversion Attempts}
We refined our original methodology for discovering triangular arbitrage sequences by relaxing the constraint for the third trade to be executed. We identify equal quantities in the first two trades in the same way as before. To determine the trade latency, we empirically explored different time constraints. $93\%$ of trades with equal quantities have a latency between $10$ms and $100$ms. Latency lower than $10$ms gives poor recall, with less than $5\%$ of total attempts. Latency greater than $100$ms only accounts for $2\%$ of all attempts. The average latency is $22$ms with a standard deviation of $15$ms. We find that, as before, $\Delta t \approx 50$ms is an approximation that likely provides both high precision and recall for identifying indirect conversion trading sequences. However, we lack ground truth to definitively evaluate this part of our methodology.

\subsubsection{Determining Profitability of Internal Conversions} The first two trades of a triangular arbitrage sequence, $c_1 \mapsto c_2 \mapsto c_3 \mapsto c_1$ are $c_1 \mapsto c_2$ and $c_2 \mapsto c_3$. Executing these two trades gives an indirect exchange ratio between $c_1$ and $c_3$. If this exchange ratio, net of fees and residue, is greater than the exchange ratio of $c_1 \mapsto c_3$, then this conversion offers a favorable exchange ratio to the direct one. We call such favorable conversions \textit{indirect internal conversions}. Since the exchange ratio between $c_1$ and $c_3$ fluctuates, it is important to define the time it is taken. We wish to approximate the trader's view of the market upon executing the internal conversion. Therefore, we take the exchange ratio at a short period of time prior to the first trade's execution time. As discussed above, $50$ms is a good delay approximation for the trader's view. Kaiko's order book data is given on a $60$-second basis, so it is unlikely that this $50$ms time window falls within Kaiko's data intervals. To approximate the order book's best bid/ask price at the time, we take the volume weighted average price (VWAP) \cite{vwap} of $c_1 \mapsto c_3$, over the period of $50$ms, prior to the execution time of the first trade. These trades tell us the best bid/ask level at that time, and taking an average, weighted by volume, is a good approximation of the trader's view of the order book. However, it is possible that within that time period, other participants posted bids and asks and then cancelled them, giving the trader a completely different view of $c_1 \mapsto c_3$. In our analysis, we assume that this did not occur. If there were no trades during that period and order book data is unavailable, we cannot determine if the conversion is profitable and did not include such conversions in our profitability analysis.

We found $26,603,038$ indirect conversions, which is $2.71\%$ of the total number of trades. $95\%$ of the indirect conversions resulted in a favorable exchange ratio. We hypothesized that unprofitable conversions occur when multiple traders obtain the same intermediary coin $c_2$, and simultaneously attempt to convert it to $c_3$. For example, one trader is looking to convert $c_1 \mapsto c_2 \mapsto c_3$ for a favorable exchange ratio to $c_1 \mapsto c_3$, and a second trader is looking to convert $y \mapsto c_2 \mapsto c_3$ at a favorable exchange ratio to $y \mapsto c_3$ ($y$ could be the same coin as $c_1$). When both traders obtain $c_2$, there could be a scenario where only one trader is able to convert $c_2$ to $c_3$ at the best bid/ask level. This happens when the first one to execute $c_2 \mapsto c_3$ consumes the entire quantity of the best bid/ask and causes the order book to change. The laggard in this case engages in a loss mitigating strategy. One option is to convert $c_2$ to $c_3$ at a worse exchange ratio than originally intended, potentially resulting in a conversion that is worse than the direct one. Another option is to convert $c_2$ to several other coins, with the goal of minimizing the potential losses from unfavorable exchange ratios. We call the former strategy a \textit{full-exit strategy} and the latter \textit{partial-exit strategy}.

To corroborate our hypothesis with the trade data, we refine our methodology to cluster competing conversions and identify loss-mitigating exit strategies.

\subsection{Clustering Competing Indirect Internal Conversion Attempts}
Using our methodology to discover indirect conversions, we identified conversions that were initiated around the same time and had an overlapping intermediary coin. This is because competing conversions try to complete the same second trade. Therefore, for a given second trade $c_2 \mapsto c_3$, we look at indirect conversions attempts of the form $x \mapsto c_2 \mapsto c_3$ that started within $100$ms of each other.

Every cluster of competing conversions can have \textit{winning conversions}, i.e., indirect conversions that were completed at a favorable rate to the direct rate and \textit{losing conversions}, conversions that completed at an unfavorable rate to the direct rate. 

Losing conversions can have many causes, such as mistiming or an inaccurate view of the order book. We believe that one of those reasons is that traders who successfully completed the first trade, and failed to complete the second trade, are unloading their $c_2$ coin at an unfavorable rate to avoid having directional exposure to $c_2$.

\subsubsection{Identifying Loss-Mitigating Trading Strategies}
For losing conversions, we wanted to see if the loss was the result of a competing winning conversion that utilized all the capacity of $c_2 \mapsto c_3$. Determining whether a conversion utilized all capacity is impossible without knowing the order book at that time. However, we can approximate it by observing the next trade of $c_2 \mapsto c_3$. Since we have the \textit{trade id}, which is monotonically increasing with the trades, we can tell whether the next trade of $c_2 \mapsto c_3$ was completed at the same price or not. If it was not completed at the same price, it is likely that the previous trade used up the capacity of the previous level. This is only a heuristic, as traders can post and cancel orders at any time.

Loss-mitigating conversions are losing conversions where a winning conversion in the same cluster used up the intermediary's coin capacity.

By analyzing the trade data, we found that $17.7\%$ of all losing conversions corresponded to loss-mitigating traders.

We identified two sub-strategies of loss-mitigating traders.

\subsubsection{Full-Exit Loss Mitigating Strategy} These are conversions that converted an equal quantity $Q_1$ from $x\mapsto c_2$ and $c_2 \mapsto c_3$, i.e., these are traders who converted all their $c_2$ into one coin.

\subsubsection{Partial-Exit Loss Mitigating Strategy} These are conversions that converted a quantity $Q_1$ from $x\mapsto c_2$ at price $p_{12}$, but converted a lower quantity $Q_2 < Q_1$ from $c_2 \mapsto c_3$, i.e., these are traders who unloaded some, but not all of their $c_2$ into one coin. This strategy can be attributed to traders solving the following minimization problem: Find a set of $k$ coins $\{d_1, d_2, \ldots d_k\}$ having exchange ratios $\{p_1, p_2, \ldots p_k\}$, and a set of $k$ quantities $\{q_1, q_2, \ldots q_k \}$, where $\sum_{j=1}^{k} q_j = Q_1$ such that the following loss function is minimized:
\begin{align*}
    Loss = \sum_{j=1}^{k} q_j\left(\frac{p_{13}}{p_{12} p_j}\right)
\end{align*}
where $p_{13}$ is the direct exchange ratio of $x \mapsto c_3$. 
To detect partial exits, we iterate over all different combinations of trades from $c_2 \mapsto c_3$ having quantities less than $Q_1$, such that they sum exactly to $Q_1$, up to fees and residue. 

We found that $85\%$ of loss-mitigating strategies are partial exits and $15\%$ are full exits. It also turns out that solving a loss minimization problems is effective, as the average partial-exit loss is $21$bps while the full-exit average loss is $25$bps.


\section{Results}
\subsection{Volume}
We found $26,603,038$ indirect conversions, which is $2.71\%$ of total trades. We found $1,381,928$ triangular arbitrage sequences, accounting for $0.24\%$ of total trades.

\begin{figure}[ht!]
  \centering
    \includegraphics[width=.50\textwidth]{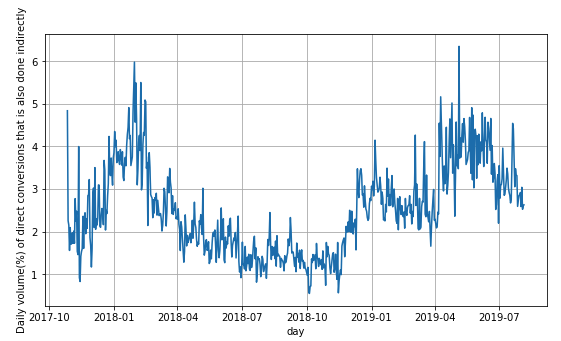}
    \caption{Percentage of daily volume of direct conversions trades that is also done indirectly. Typically, $1\%-3\%$ of the daily volume of a direct conversion pair is also executed via indirect conversion sequences.} \label{pct_of_pc_vol_chart1}
\end{figure}

\begin{figure}[ht!]
  \centering
    \includegraphics[width=.50\textwidth]{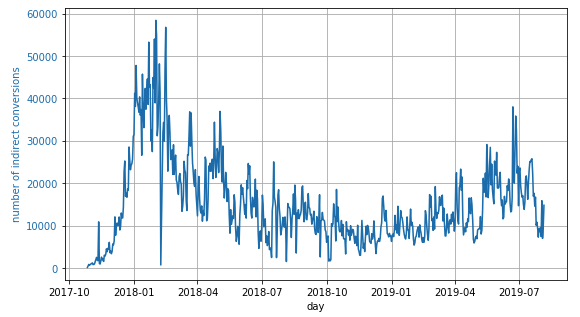}
    \caption{Daily number of indirect conversion attempts. From October $2017$ - October $2018$, the number of indirect conversions has been trending down and since October $2018$ has been trending up.} \label{num_pcs_absolute}
\end{figure}

\begin{figure}[ht!]
  \centering
    \includegraphics[width=.50\textwidth]{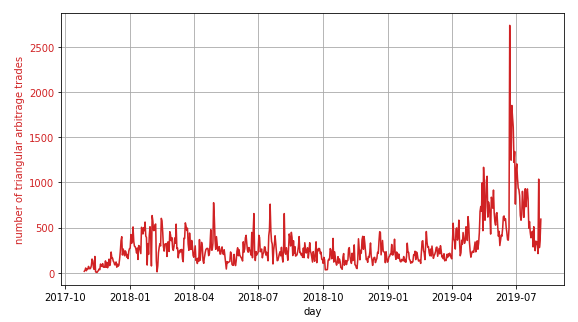}
    \caption{Daily number of triangular arbitrage attempts. From October $2017$ - April $2019$, the number of triangular arbitrage attempts has been trending down and since April $2019$ has been trending up, with a sharp spike in July $2019$} \label{num_tris_absolute}
\end{figure}

The time series of the number of favorable indirect conversion attempts as a percentage of number of direct conversions, on a daily basis, is shown in Figure \ref{pct_of_pc_vol_chart1}. The number of favorable indirect conversion attempts, on a daily basis, in shown in Figure \ref{num_pcs_absolute}. The number of triangular arbitrage attempts, on a daily basis, in shown in Figure \ref{num_tris_absolute}.

\subsection{Latency}
\begin{figure}[ht!]
  \centering
    \includegraphics[width=.24\textwidth, height=4cm]{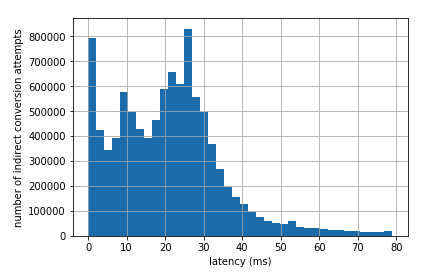}
    \includegraphics[width=.24\textwidth, height=4cm]{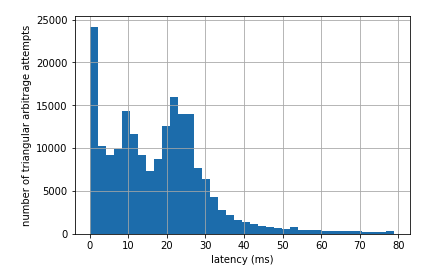}
    \caption{Left: Latency distribution, in ms, of indirect conversion trade sequences. Right: Latency distribution, in ms, of triangular arbitrage trade sequences. Indirect conversion trades exhibit higher latency with a fatter tail, indicating that triangular arbitrage is more competitive.} \label{fig:latency_dist_chart}
\end{figure}

We found that indirect conversion sequences have, on average, $21.43$ms of delay between the first trade and the second, with a standard deviation of $11.04$ms. $78.94\%$ have a latency of $30$ms or less.
Triangular arbitrage sequences are faster than indirect conversions. This could be an indication that this space is more competitive. We found that triangular arbitrage sequences have, on average, $20.7$ms of delay between consecutive trades, with a standard deviation of $11.04$ms. $85.43\%$ have a latency of $30$ms or less. The latency statistics of triangular arbitrage sequences and indirect conversions are shown in Table \ref{latency_stats_table}.
\begin{table}[ht!]
\begin{center}
    \begin{tabular}{||c|c|c||}
    \hline \text{Type} & \text{Metric} & \text{Value} \\\hline
    \text{Triangular Arbitrage} & \text{Latency Average} & $20.7$ms \\ \hline
    \text{Triangular Arbitrage} & \text{Latency Stdev} & $11.04$ms \\ \hline
    \text{Triangular Arbitrage} & \text{\% Below $30$ms} & $85.43\%$ \\ \hline
        \text{Indirect Conversion} & \text{Latency Average} & $21.43$ms \\ \hline
    \text{Indirect Conversion} & \text{Latency Stdev} & $13.7$ms \\ \hline
    \text{Indirect Conversion} & \text{\% Below $30$ms} & $78.94\%$ \\ \hline
\end{tabular}
\end{center}
\caption{Latency statistics for triangular arbitrage and indirect conversions. The triangular arbitrage strategy exhibits lower latency than indirect conversions, possibly a sign that this strategy is more competitive }\label{latency_stats_table}
\end{table}
The latency distribution, in ms, of indirect conversions and triangular arbitrage is shown in Figure \ref{fig:latency_dist_chart}.
\subsection{Profitability}
We calculate the equal-weighted return and the return on capital for triangular arbitrage trades and indirect conversions. The equal-weighted return is the average of returns. The return on capital is the arithmetic average of returns, weighted by quantity of base coin committed to the trade, i.e., larger trades receive higher weight.

Indirect conversion attempts were profitable $93.92\%$ of the time with an equal-weighted net return of $11.8$bps. The return on capital for indirect conversions is $22\%$ greater, at $14.4$bps, which means that traders efficiently commit more coins to more profitable opportunities. 

Triangular arbitrage trades were profitable $94.97\%$ of the time with an equal-weighted net return of $9.3$bps.  The return on capital for triangular arbitrage sequences is $5\%$ greater, at $9.8$bps, which is slightly higher than the equal-weighted return, but not significantly. 

It appears that triangular arbitrage traders are unable to significantly allocate more coins to more profitable opportunities. One potential explanation is that the space is more crowded, or ``arb-ed out,'' than the indirect conversions strategy.

\subsection{Loss-Mitigating Strategies}
We saw $1,617,464$ instances of indirect conversion completed at an unfavorable rate compared to the direct rate, or $6.08\%$ of the total number of indirect conversions. $17.7\%$ of such conversions were loss-mitigating trades, triggered by a competing indirect conversion that used all the intermediary coin's capacity. When this happened, there were $1.06$ competing conversions on average. The highest number of competing conversions we saw in one cluster was $5$.

In $15\%$ of such cases, the entire base quantity was unloaded into another coin, i.e., they were full exits. The return on capital for the full exit strategy was $-27.4$bps. $85\%$ of the time, the base coin quantity was unloaded to several other coins, i.e., they were partial exits. On average, the number of coins used to exit was $2.7$, and the return on capital was $-23.8$bps. Intuitively, this shows that solving a loss minimization problem mitigates losses by $3.6$bps compared to a full exit. Note, however, that a full exit could potentially be the solution to the minimization problem as well.

\section{Discussion}
The existence of triangular arbitrage in traditional foreign exchange markets is well documented and studied in academia. In centralized cryptocurrency exchanges such as Binance, triangular arbitrage happens at a small scale. However, we found that a different strategy is taking place, which converts two coins through an intermediary coin, at a favorable rate to the direct conversion rate. This strategy comes with a risk, however, that multiple competing conversions occur at the same time, preventing slower ones from completing their conversions. We saw that when this happens, traders engage in a loss-mitigating strategy. We believe that the rationale for this strategy stems from Binance's fee structure, where there is a fee for market-making, thus creating frictions for the triangular arbitrage strategy. We believe arbitrageurs adapted to the fee structure by executing the indirect conversions strategy. While triangular arbitrage increases the arbitrageur's holding in the base asset, indirect conversions simply provide a more favorable ratio to the direct one. Participants who already have a stake in the cryptocurrency ecosystem and wish to rebalance their portfolio might choose to engage in this strategy.

\section{Conclusion}
We found that $0.24\%$ of daily trades on Binance can be attributed to triangular arbitrage trades. We found that a different strategy is $11$ times more prevalent, at $2.71\%$ of daily trades, which involves exchanging coins through an intermediary coin at a favorable rate to the direct exchange rate. We designed a methodology to measure such events and discovered that $93.92\%$ of the time the rate obtained from these conversions is $14.4$bps better than the direct one. When it is not, $17.7\%$ of the time traders engage in loss-mitigating strategies, and we identified two of them.

\section*{Acknowledgements}
The authors would like to thank Andrew Papanicolaou and David Yermack  for their feedback and suggestions to improve the quality of our manuscript. We acknowledge funding support under National Science Foundation award 1844753. 




\bibliographystyle{abbrv}
\bibliography{refs.bib}

\section{Appendix}
In this appendix, we define notations that are specific to the market microstructure of Binance and formally define triangular arbitrage and indirect conversions. 

Let $\mathbb{P}$ be the set of all \textit{pairs} $x/y$ traded on Binance. Every pair facilitates two conversions; from $x$ to $y$ and from $y$ to $x$, denoted by $(x \mapsto y)$ and $(y \mapsto x)$, respectively. 

\begin{definition} \textbf{(exchange ratio)} Let $\psi_{ij}$ be the proceeds of converting 1 unit of coin $c_i$ to $c_j$. We call $\psi_{ij}$ the \textit{exchange ratio} and it is given by
\begin{align*}
    \psi_{ij} = 
    \begin{cases}
        \frac{1}{p_{c_j/c_i}^{ask}} & \text{if $c_j/c_i \in \mathbb{P}$} \\
        p_{c_i/c_j}^{bid} & \text{if $c_i/c_j \in \mathbb{P}$} \\
    \end{cases}
\end{align*}
\end{definition}
\begin{definition} \textbf{(pair capacity)} Denote $\eta_{ij}$ the maximum quantity $c_i$ that can be converted to $c_j$. We call $\eta_{ij}$ the \textit{capacity} of $(c_i \mapsto c_j)$ and it is given by
\begin{align*}
    \eta_{ij} = 
    \begin{cases}
        p_{c_j/c_i}^{ask}q_{c_j/c_i}^{ask} & \text{if $c_j/c_i \in \mathbb{P}$} \\
        q_{c_i/c_j}^{bid} & \text{if $c_i/c_j \in \mathbb{P}$} \\
    \end{cases}
\end{align*}
\end{definition}
\begin{definition} \textbf{(bid-ask spread)} We write
\begin{align*}
\Delta_{ij} = 
    \begin{cases}
        p_{c_j/c_i}^{ask}-p_{c_j/c_i}^{bid} & \text{if $c_j/c_i \in \mathbb{P}$} \\
        p_{c_i/c_j}^{ask}-p_{c_i/c_j}^{bid} & \text{if $c_i/c_j \in \mathbb{P}$} \\
    \end{cases}
\end{align*}
the bid ask spread of pair $(c_i \mapsto c_j)$
\end{definition}
\begin{definition} \textbf{(trade quantity)} suppose we wish to convert $q$ units of $c_i$ to $c_j$. The \textit{trade quantity} passed to the exchange as a parameter is denoted $T(q)$ and is given by
\begin{align*}
    T_{ij}(q) = \begin{cases}
        \frac{q}{p_{c_j/c_i}^{ask}}  & \text{if $c_j/c_i \in \mathbb{P}$} \\
        q & \text{if $c_i/c_j \in \mathbb{P}$} \\
    \end{cases}
\end{align*}
\end{definition}
\begin{definition} \textbf{(minimum trade lot size)} Let $m_{ij}$ be the minimum amount of $c_i$ that must be converted to $c_j$ in $\pair{$c_i$}{$c_j$}$, i.e., when converting $q$ units of $c_i$ to $c_j$, $m_{ij}\leq q$
\end{definition}
\begin{definition} \textbf{(high-level parameters)} the $h$-th order book exchange ratio and capacity are denoted $\psi_{ij}^{h}$ and $\eta_{ij}^{h}$, respectively.
\end{definition}
\begin{definition} \textbf{(minimum price increment)}
Denote $dx_{ij}$ the the minimum order book price increments of $(c_i \mapsto c_j)$. 
\end{definition}
\begin{remark*}
It holds that $dx_{ij} = dx_{ji}$.
\end{remark*}
In practice, when converting $q$ units of $c_i$ to $c_j$, the trade quantity passed to the exchange as a parameter has to be a multiple of $dx_{ij}$, therefore only $\left \lfloor \frac{T(q)}{dx_{ij}} \right \rfloor dx_{ij}$
units are passed. Denote $\left \lfloor x \right \rfloor _{y} \eqdef \left \lfloor \frac{x}{y} \right \rfloor y $ the rounding operation of $x$ to the nearest multiple of $y$. 
\begin{definition} \textbf{(residuals)} Let $r_{ij}(q)$ be the quantity of $c_i$ left when converting $q$ units of $c_i$ to $c_j$. We write
\begin{align*}
    r_{ij}(q) = \begin{cases}
        \left(T(q) - \left\lfloor T(q)\right\rfloor _{dx_{ij}} \right)p_{c_j/c_i}^{ask}   & \text{if $c_j/c_i \in \mathbb{P}$} \\
        T(q) - \left\lfloor T(q)\right\rfloor _{dx_{ij}} & \text{if $c_i/c_j \in \mathbb{P}$} \\
    \end{cases}
\end{align*}
\end{definition}
Therefore, converting $x$ units of $c_i$ yields $\psi_{ij}\left[x-r_{ij}(x)\right]$ units of $c_j$ and $r_{ij}(x)$  \textit{residual} units of $c_i$
\subsection{Cycle Arbitrage}
The set of pairs  $C=\{(c_1 \mapsto c_2), (c_2 \mapsto c_3), \ldots, (c_n \mapsto c_{n+1})\} $ is called a \textit{cycle}, if $c_{n+1}=c_1$ and $c_j \not \in \{c_1,\ldots\,c_{j-1}\}$ for $1<j\leq n$. In the context of cycles, we use the shortened notation $\psi_i, \eta_{i}, dx_i, r_i$ when referring to pairs of the form $(c_i \mapsto c_{i+1})$. A cycle with $n=3$ is called a \textit{triangular sequence}.
\begin{definition}
The \textit{capacity} of cycle $C=\{(c_1\mapsto c_2), \ldots, (c_n \mapsto c_{n+1})\}$ is the maximum quantity of $c_1$ that can be converted through its pairs.
\end{definition}
The capacity of a cycle is given by
\begin{align}\label{eq:Q}
Q = \min_{1 \leq i \leq n} \left\{ \eta_{i}\left( \prod_{k=1}^{i-1}\psi_{k}\right)^{-1} \right\}
\end{align}
The balance of $c_i$, denoted $q_i$, is given by the recurrence relation
\begin{align*}
    q_1 &= Q - r_1(Q) \\
    q_{i+1} &= \psi_i\left[q_i - r_i(q_i)\right], \qquad 1 \leq i \leq n
\end{align*}

\begin{remark*}
$r_i\Big(x-r_i(x)\Big) = 0$ 
\end{remark*}
\subsection{Binance Fee Structure}
The fees paid on $(c_i \mapsto c_{i+1})$ are $fq_{i+1}b_{i+1}$ where $f=5\cdot10^{-4}$
and $b_{i+1}$ is the last traded price of $(c_{i+1} \mapsto \text{BNB})$
\begin{align*}
    b_{i} = 
    \begin{cases}
        \frac{1}{p_{\text{BNB}/c_i}} & \text{if $\text{BNB}/c_i \in \mathbb{P}$} \\
        p_{c_i/\text{BNB}} & \text{if $c_i/\text{BNB} \in \mathbb{P}$} \\
    \end{cases}
\end{align*}
The gain/loss in $c_1$ terms is given by
\begin{align}\label{eq:gain function}
    G =(q_{n+1}-q_{1}) + \underbrace{\sum_{i=1}^{n}r_i(q_i)\psi_{i1}}_{\text{residuals $c_1$ value}} - \underbrace{f\sum_{i=1}^{n}q_{i+1} b_{i+1}\psi_{B1}}_{\text{fees $c_1$ value}}
\end{align}
Note that $G$ is a function of $\{\psi_{21}, \ldots, \psi_{n1},\psi_{B1},b_2, \ldots, b_{n+1}\}$ as well as $\psi_1, \ldots \psi_n$. 
\begin{definition}
a cycle is called an \textbf{arbitrage free cycle} if $G \leq 0$
\end{definition}
\begin{definition}
a cycle is called an \textbf{open cycle} if $G>0$
\end{definition}
\begin{remark*}
One can argue that there are additional costs for converting the residuals to $c_1$ and purchasing BNB tokens ahead of time to be able to pay fees, which are not accounted for in
\eqref{eq:gain function}. However, this can be done in infrequent bulk trades. We make the assumption that small directional exposure to BNB and residuals is negligible compared to the accumulated gains over a period of time.
\end{remark*}

\begin{remark} 
If we assume zero residuals, i.e., $\forall  i : r_i = 0,$ then $q_{i+1} = Q\prod_{j=1}^{i-1}\psi_j$ and \eqref{eq:gain function} reduces to $G = Q\left(\prod_{i=1}^{n}\psi_i - 1\right)-f\sum_{i=1}^{n}Q\left(\prod_{j=1}^{i}\psi_{j}\right)b_{i+1}\psi_{B1}$
\end{remark}

\subsection{Indirect Internal Conversions} 
The set of pairs $V = \{(c_1\mapsto c_2), (c_2 \mapsto c_3), \ldots, (c_n \mapsto c_{n+1})\}$ 
is called a \textit{conversion} if $c_j \not \in \{c_1,\ldots\,c_{j-1}\}$ for $1<j\leq n+1$. In the context of conversions, we use
notation $\psi_i, \eta_i, dx_i, r_i$ when referring to pairs of the form $(c_i \mapsto c_{i+1})$. When the intermediate pairs are clear, we use the shortened notation $c_1 \stackrel{V}{\leadsto} c_{n+1}$  
\begin{definition}
The \textit{capacity} of conversion $V=\{(c_1\mapsto c_2), \ldots, (c_n \mapsto c_{n+1})\}$ is the maximum quantity of $c_1$ that can be converted through its pairs and is defined similarly to the capacity of a cycle.
\end{definition}
\begin{definition} \textbf{(conversion proceeds)} Let $c_1 \stackrel{V}{\leadsto} c_{n+1}$  be a conversion from $c_1$ to $c_{n+1}$ with capacity $Q$. The conversion proceeds of $q\leq Q$ units of $c_1$ is the quantity of $c_{n+1}$ that results from converting through the pairs of $V$, accounting for residuals and fees, i.e.,
\begin{align*}
    Pr(q,V) = q_{n+1} +
    \underbrace{\sum_{i=1}^{n}r_i(q_i)\psi_{i(n+1)}}_{\text{residuals $c_{n+1}$ value}} -\underbrace{f\sum_{i=1}^{n}q_{i+1}b_{i+1}\psi_{B(n+1)} }_{\text{fees $c_{n+1}$ value}}
\end{align*}

\end{definition}
\begin{definition} \textbf{(profitable conversions)}  Let $x \stackrel{V_1}{\leadsto} y$ and $x \stackrel{V_2}{\leadsto} y$ be two conversions from $x$ to $y$, with capacities $Q_1$ and $Q_2$, respectively. Let $Q=\min\{Q_1,Q_2\}$. We say that $V_1$ is a profitable conversion w.r.t.\ $V_2$ if the proceeds of conversion $V_1$ are greater than the proceeds of conversion $V_2$, i.e.,  $Pr(Q,V_1) > Pr(Q,V_2)$
\end{definition}
\subsection{Binance Order Types}
Binance supports the following order types:
\begin{enumerate}
    \item \texttt{LIMIT} - an order to buy/sell a pair at a specified price and can only be executed at that price (or better). Not guaranteed to execute
    \item \texttt{MARKET} - an order to buy/sell a pair at the best current market price, i.e., lowest ask or highest bid.
    \item \texttt{STOP\_LOSS\_LIMIT} - an order to buy (sell) a pair, once its price exceeds (drops below) the specified price. In contrast to a \texttt{LIMIT} order, the price should be above (below) the lowest ask (highest bid). The execution price is guaranteed to be the specified price.
    \item \texttt{STOP\_LOSS} - same as \texttt{STOP\_LOSS\_LIMIT}, but when the price threshold is breached, a \texttt{MARKET} order is executed and the execution price is not guaranteed.
    \item \texttt{TAKE\_PROFIT\_LIMIT} - equivalent to a \texttt{LIMIT} order
    \item \texttt{TAKE\_PROFIT} - automatically places a \texttt{MARKET} order when the specified price level is met
    \item \texttt{LIMIT\_MAKER} - \texttt{LIMIT} orders that will be rejected if they would immediately match and trade as a taker.
    \item \texttt{ICEBERG} - an order used for large quantities as it automatically breaks down to multiple \texttt{LIMIT} orders with different prices. The goal is to hide the actual order quantity and prevent trend-followers from unfavorably moving the price.
\end{enumerate}

\end{document}